**J. Tõke et al. reply:** The Comment does not challenge any of the conclusions of the targeted paper [1], neither factually nor formally. Instead, it attacks a claim that has never been made in Ref. [1]. In fact, one can verify that the Comment is consistent with the statement that all conclusions of ref. [1] are warranted, based on the full set of observations discussed in that paper.

In the body and summary, the Comment merely demonstrates, that a certain subset [listed in the Comment as (1) and (2)] of all observations used to sustain the conclusions made in Ref. [1], provides insufficient evidence for these conclusions. Such demonstration bares no relevance to the validity of the arguments made in Ref. [1]. Clearly, a statement, "A and B imply C", is not invalidated or challenged by the observation that B alone does not imply C. Therefore, the fact that a statistical model can account for the simultaneous effects of a saturation of light-particle multiplicities (and transverse energies) and a rise of the IMF transverse energy, with increasing IMF multiplicity does not contradict the conclusions of Ref. [1] based on a larger set of observations.

We emphasize that the conclusions made in Ref. [1] are based on more than just the subset [(1) and (2)] of observations selected in the Comment. Among these observations, but not included in the subset considered in the Comment, are (i) the emission patterns of intermediate-mass fragments (IMF), (ii) the strength of the auto-correlation between IMF multiplicity $m_{IMF}$ and total transverse energy $E^t_{tot}$, and (iii) the character of the correlation between $m_{IMF}$ and the velocity or energy of coincident projectile-like fragments. Notably, Phair et al. find themselves unable to account for the above crucial facts. It is argued in Ref. [1] that together with (i) – (iii), and not separately, the two listed observations (1) and (2) are inconsistent with a statistical IMF production mechanism. The Comment does not challenge this claim.

Within the authors' choice of words, the Comment is consistent with Ref. [1] and is at the same time intrinsically correct. Expressions such as "inferred partly" (opening paragraph) and "helps prove" (paragraph before last) confirm that only a subset of observations is considered in the Comment. Similarly, the conclusions of the Comment explicitly refer to only the selected subset [(1) and (2)] of the full set discussed in Ref. [1] and not to the full set itself.

While inconsequential for the conclusions of either Comment or Ref. [1], there are several misstatements of facts in the Comment worth pointing out:

1) Contrary to what is stated in the opening paragraph of the Comment, no allegation can be found in Ref. [1] that the "saturation" feature is inconsistent with statistical models. On the contrary, in Ref. [1] this feature was interpreted in terms of statistical emission and taken to indicate saturation in total excitation energy $E^*$. This is the same interpretation as presented now in the Comment. Accordingly, an equilibrium-statistical code (EVAP) was used in Ref. [1] to extract the corresponding value of the limiting $E^*$.

2) Contrary to what is stated in the second paragraph of the Comment, Ref. [1] does not allege that, taken alone, the two observations listed as (1) and (2) in the Comment prove dynamical behavior. The logics of Ref. [1] suggests that neither this nor any other subset of the observations discussed in Ref. [1] is sufficient to conclude a dynamical IMF emission scenario.

3) By failing to refer to prior works, the Comment creates the impression that the authors of Ref. [1] must have been unaware of the significance of auto-correlations between $m_{IMF}$ and the IMF transverse energy $E^t_{IMF}$. In fact, it were the authors of Ref. [1] who have claimed [2] in the past the importance of such autocorrelations, while the authors of the Comment have consistently rejected [3,4] these claims. Moreover, in their published papers [3,4], the authors of the Comment have relied critically on their assertion that, for fixed total excitation energy $E^*$, $m_{IMF}$ is not correlated with $E^t_{tot}$ in a thermal emission scenario. The latter is now contradicted by Fig. 1, bottom panel in the Comment. In this figure, for the range of $m_{IMF}$ corresponding to saturation in light-product multiplicity and, hence, for a fixed excitation energy (according to the Comment), the total transverse energy shows a definite, albeit much weaker than observed experimentally, correlation with IMF multiplicity, $\Delta E^t_{tot}/\Delta m_{IMF} \approx 15$ MeV/IMF.

Lastly, to answer the question posed in the title of the Comment, the observations listed in the Comment are not direct evidence for dynamical fragment production. However, it appears that the full set of observations discussed in Ref. [1] is incompatible with thermal emission scenarios and, hence, does favor a dynamical scenario [1].

This work has been supported by the U.S. Department of Energy Grant. No. DE-FG02-88ER-40414.

J.Tõke and W.U. Schröder
*Department of Chemistry and Nuclear Structure Research Laboratory, University of Rochester, Rochester, New York 14627*

[1] J. Tõke et al., Phys. Rev. Lett. **77**, 3514 (1996).
[2] W. Skulski et al., DOE Progress Report DOE/ER/40414-8, p.75, Rochester 1995, and in Proceedings of XIII Winter Workshop on Nuclear Dynamics, Marathon 1997.
[3] L.G. Moretto, R. Ghetti, L. Phair, K. Tso, and G.J. Wozniak, LBNL-39388 Preprint, p. 23, (September 1996). Published in Phys. Rep. **287**, 249 (1997).
[4] R. Ghetti, L.G. Moretto, L. Phair, K. Tso, G.J. Wozniak, LBNL 39196 preprint, p. 23, (December 1996). Submitted to Nucl. Phys.